\documentclass{article}
\usepackage[T1]{fontenc}
\usepackage[latin1]{inputenc}
\usepackage{a4wide}
\makeatletter

\newcommand{\LyX}{L\kern-.1667em\lower.25em\hbox{Y}\kern-.125emX\@}

\newcommand{\lyxaddress}[1]{
  \par {\raggedright #1 
  \vspace{1.4em}
  \noindent\par}
}

\makeatother

\begin{document}

\title{Mixedness in Bell-violation vs. Entanglement of Formation }

\author{Sibasish Ghosh\protect\( ^{\%}\protect \)\thanks{
res9603@isical.ac.in
} , Guruprasad Kar\protect\( ^{\%}\protect \)\thanks{
gkar@isical.ac.in
} , Aditi Sen(De)\protect\( ^{\#}\protect \)\thanks{
dhom@boseinst.ernet.in
} and Ujjwal Sen\protect\( ^{\#}\protect \)\protect\( ^{\ddagger }\protect \)}

\maketitle

\lyxaddress{\protect\( ^{\%}\protect \)Physics and Applied Mathematics Unit, Indian Statistical
Institute, 203 B.T. Road, Kolkata 700 035, India }

\lyxaddress{\protect\( ^{\#}\protect \)Department of Physics, Bose Institute, 93/1 A.P.C.
Road, Kolkata 700 009, India}

\begin{abstract}
Recently Munro, Nemoto and White (\emph{The Bell Inequality: A measure of Entanglement?},
quant-ph/0102119) tried to indicate that the reason behind a state \( \rho  \)
having higher amount of entanglement (as quantified by the entanglement of formation)
than a state \( \rho ^{\prime } \), but producing the same amount of Bell-violation,
is due to the fact that the amount of mixedness (as quantified by the linearised
entropy) in \( \rho  \) is higher than that in \( \rho ^{\prime } \). We counter
their argument with examples. We extend these considerations to the von Neumann
entropy. Our results suggest that the reason as to why equal amount of Bell-violation
requires different amounts of entanglement cannot, at least, be explained by
mixedness alone.
\end{abstract}
Werner\cite{1} (see also Popescu\cite{2}) first demonstrated the existence
of states which are entangled but do not violate any Bell-type inequality\cite{3,4}.
But there exist classes of states (pure states, mixture of two Bell states),
which violate Bell inequality whenever they are entangled\cite{5,6}.

This implies that to produce an equal amount of Bell-violation, some states
require to have more entanglement (with respect to some measure) than others.
It would be interesting to find out what property of the first state requires
it to have more entanglement to produce the same Bell-violation. Recently Munro
\emph{et al.}\cite{7} have tried to indicate that this anomalous property of
the first state is due to its being more \emph{mixed} than the second, where
they took the linearised entropy\cite{8} as the measure of mixedness. 

As in \cite{7}, we use the entanglement of formation as our measure of entanglement.
For a state \( \rho  \) of two qubits, its entanglement of formation \( EoF(\rho ) \)
is given by\cite{9} 
\[
EoF(\rho )=h\left( \frac{1+\sqrt{1-\tau }}{2}\right) \]
with
\[
h(x)=-x\log _{2}x-(1-x)\log _{2}(1-x).\]
The tangle \( \tau  \) \cite{10} is given by
\[
\tau (\rho )=[\max \{0,\: \lambda _{1}-\lambda _{2}-\lambda _{3}-\lambda _{4}\}]^{2},\]
the \( \lambda _{i} \)'s being square root of eigen values, in decreasing order,
of \( \rho \widetilde{\rho } \), where
\[
\widetilde{\rho }=(\sigma _{y}\otimes \sigma _{y})\rho ^{*}(\sigma _{y}\otimes \sigma _{y}),\]
the complex conjugation being taken in the standard product basis \( \left| 00\right\rangle  \),
\( \left| 01\right\rangle  \), \( \left| 10\right\rangle  \), \( \left| 11\right\rangle  \)
of two qubits. Note that EoF is monotonically increasing ranging from \( 0 \)
to \( 1 \) as \( \tau  \) increases from \( 0 \) to \( 1 \) and hence, like
Munro \emph{et al.}\cite{7}, we take \( \tau  \) as our measure of entanglement. 

The maximum amount of Bell-violation(\( B \)) of a state \( \rho  \) of two
qubits is given by\cite{6} 

\[
B(\rho )=2\sqrt{M(\rho )}\]
where \( M(\rho ) \) is the sum of the two larger eigenvalues of \( T_{\rho }T^{\dagger }_{\rho } \),
\( T_{\rho } \) being the \( 3\times 3 \) matrix whose \( (m,n) \)-element
is
\[
t_{mn}=tr(\rho \sigma _{n}\otimes \sigma _{m}).\]

The \( \sigma  \)'s are the Pauli matrices. 

The linearised entropy \cite{8}

\[
S_{L}(\rho )=\frac{4}{3}(1-tr(\rho ^{2}))\]
is taken as the measure of mixedness.

Munro \emph{et al.}\cite{7} proposed that given two two-qubit states \( \rho  \)
and \( \rho ^{\prime } \) with
\[
B(\rho )=B(\rho ^{\prime }),\]
 but
\[
\tau (\rho )>\tau (\rho ^{\prime }),\]
would imply 
\[
S_{L}(\rho )>S_{L}(\rho ^{\prime }).\]
To support this proposal, it was shown that it holds for any combination of
states from the following three classes of states: 

(1) the class of all pure states
\[
\rho _{pure}=P[a\left| 00\right\rangle +b\left| 11\right\rangle ]\]
with \( a,\: b\geq 0 \),and \( a^{2}+b^{2}=1, \) 

(2) the class of all Werner states\cite{1}
\[
\rho _{werner}=xP[\Phi ^{+}]+\frac{1-x}{4}I_{2}\otimes I_{2}\]
with \( 0\leq x\leq 1 \) and \( \Phi ^{+}=\frac{1}{\sqrt{2}}(\left| 00\right\rangle +\left| 11\right\rangle ) \),
and 

(3) the class of all maximally entangled mixed states\cite{11}
\[
\rho _{mems}=\frac{1}{2}(2g(\gamma )+\gamma )P[\Phi ^{+}]+\frac{1}{2}(2g(\gamma )-\gamma )P[\Phi ^{-}]+(1-2g(\gamma ))P[\left| 01\right\rangle \left\langle 01\right| \]
with \( g(\gamma )=1/3 \) for \( 0<\gamma <2/3 \) and \( g(\gamma )=\gamma /2 \)
for \( 2/3\leq \gamma \leq 1 \), and \( \Phi ^{\pm }=\frac{1}{\sqrt{2}}(\left| 00\right\rangle \pm \left| 11\right\rangle ) \).

However, consider the class of all mixtures of two Bell states
\[
\rho _{2}=wP[\Phi ^{+}]+(1-w)P[\Phi ^{-}],\]
with \( 0<w<1 \). \( \rho _{2} \) is entangled whenever \( w\neq \frac{1}{2} \),
and for that entire region, \( \rho _{2} \) is Bell-violating\cite{6}. For
this class it is easy to show that
\[
B=2\sqrt{1+\tau }\]

But the corresponding curve for pure states \( \rho _{pure} \) is also given
by\cite{7}
\[
B=2\sqrt{1+\tau }\]
We see that for any fixed Bell-violation, the corresponding \( \rho _{2} \)
has its tangle equal to that for the corresponding pure state. But the mixedness
of \( \rho _{2} \) is obviously \emph{larger} than that of the pure state (as
the mixedness is always zero for pure states).

Next consider the following class of mixtures of \emph{three} Bell states 
\[
\rho _{3}=w_{1}P[\Phi ^{+}]+w_{2}P[\Phi ^{-}]+w_{3}P[\Psi ^{+}]\]
with \( 1\geq w_{1}\geq w_{2}\geq w_{3}\geq 0 \), \( \sum _{i}w_{i}=1 \) and
\( \Psi ^{+}=\frac{1}{\sqrt{2}}(\left| 01\right\rangle +\left| 10\right\rangle ) \).
We take \( w_{1}>\frac{1}{2} \) so that \( \rho _{3} \) is entangled \cite{12}. 

For \( \rho _{3} \), we have (as \( w_{1}\geq w_{2}\geq w_{3} \))

\[
B(\rho _{3})=2\sqrt{2-4w_{2}(1-w_{2})-4w_{3}(1-w_{3})},\]
\[
\tau (\rho _{3})=1-4w_{1}(1-w_{1}),\]
\[
S_{L}(\rho _{3})=\frac{4}{3}\{w_{1}(1-w_{1})+w_{2}(1-w_{2})+w_{3}(1-w_{3})\}.\]

Let
\[
\rho ^{\prime }_{3}=w^{\prime }_{1}P[\Phi ^{+}]+w_{2}^{\prime }P[\Phi ^{-}]+w^{\prime }_{3}P[\Psi ^{+}]\]
with \( 1\geq w^{\prime }_{1}\geq w^{\prime }_{2}\geq w^{\prime }_{3}\geq 0 \),
\( \sum _{i}w^{\prime }_{i}=1 \), \( w^{\prime }_{1}>\frac{1}{2} \) be such
that 
\[
B(\rho _{3})=B(\rho _{3}^{\prime })\]
which gives
\[
w_{2}(1-w_{2})+w_{3}(1-w_{3})=w_{2}^{\prime }(1-w^{\prime }_{2})+w^{\prime }_{3}(1-w^{\prime }_{3}).\]
Now if
\[
\tau (\rho _{3})>\tau (\rho _{3}^{\prime }),\]
we have
\[
w_{1}(1-w_{1})<w^{\prime }_{1}(1-w^{\prime }_{1})\]
so that
\[
w_{1}(1-w_{1})+w_{2}(1-w_{2})+w_{3}(1-w_{3})<w^{\prime }_{1}(1-w^{\prime }_{1})+w^{\prime }_{2}(1-w^{\prime }_{2})+w_{3}^{\prime }(1-w_{3}^{\prime })\]
that is
\[
S_{L}(\rho _{3})<S_{L}(\rho _{3}^{\prime }).\]
Thus for a fixed Bell-violation, the order of \( S_{L} \) for \( \rho _{3} \)
and \( \rho _{3}^{\prime } \) is \emph{always} reversed with respect to the
order of their \( \tau  \)'s. That is, the indication of \cite{7}, referred
to earlier, is \emph{always} violated for any two states from the class of mixtures
of \emph{three} Bell states. 

One can now feel that if the \emph{entanglement of formation of two states are
equal}, it could imply some order between the amount of Bell-violation and mixedness
of the two states. But even that is not true. 

For our first example, if
\[
\tau (\rho _{2})=\tau (\rho _{pure})\]
then
\[
B(\rho _{2})=B(\rho _{pure}),\]
but 
\[
S_{L}(\rho _{2})>S_{L}(\rho _{pure}).\]

On the other hand for our second example, if
\[
\tau (\rho _{3})=\tau (\rho ^{\prime }_{3})\]
then
\[
B(\rho _{3})>B(\rho ^{\prime }_{3})\]
implies
\[
S_{L}(\rho _{3})<S_{L}(\rho ^{\prime }_{3}).\]

In Ref.\cite{7}, the linearised entropy was the only measure of mixedness that
was considered. But the von Neumann entropy\cite{13} 
\[
S(\rho )=-tr(\rho log_{4}\rho ),\]
of a state \( \rho  \) of two qubits, is a more physical measure of mixedness
than the linearised entropy. We have taken the logarithm to the base \( 4 \)
to normalise the von Neumann entropy of the maximally mixed state \( \frac{1}{2}I_{2}\otimes \frac{1}{2}I_{2} \)
to unity as it is for the linearised entropy. One may now feel that the conjecture
under discussion may turn out to be true if we change our measure of mixedness
from linearised entropy to von Neumann entropy. 

But both the von Neumann entropy and the linearised entropy are convex functions,
attaining their maximum for the same state \( \frac{1}{2}I_{2}\otimes \frac{1}{2}I_{2} \)
and each of them are symmetric about the maximum. Thus
\[
S_{L}(\rho )>S_{L}(\rho ^{\prime })\]
would imply
\[
S(\rho )>S(\rho ^{\prime })\]
and viceversa. Thus all our considerations with linearised entropy as the measure
of mixedness would carry over to von Neumann entropy as the measure of mixedness.

Our results emphasize that the reason as to why equal amount of Bell-violation
requires different amounts of entanglement cannot, at least, be explained by
mixedness alone.

We thank Anirban Roy and Debasis Sarkar for helpful discussions. We acknowledge
Frank Verstraete for encouraging us to carry over our considerations to the
von Neumann entropy. A.S. and U.S. thanks Dipankar Home for encouragement and
U.S. acknowledges partial support by the Council of Scientific and Industrial
Research, Government of India, New Delhi.

\end{document}